\def\n{\noindent}

\def\P{$P$}

\def\eq{\enskip =\enskip}
\def\pls{\enskip +\enskip}
\def\mns{\enskip -\enskip}

  \def\ket{\vert \vert  \{ \emptyset \} \rangle}
  \def\ket2{\vert \vert \otimes \{ R \} \rangle}

\def\.#1{\mathaccent 95#1}
\def\^#1{\mathaccent 94 #1}
\def\~#1{\mathaccent "7E #1}

\def\plus{\enskip +\enskip}

\def\eq{\enskip =\enskip}
\def\pls{\enskip +\enskip}
\def\mns{\enskip -\enskip}

  \def\ket{\vert \vert  \{ \emptyset \} \rangle}
  \def\ket2{\vert \vert \otimes \{ R \} \rangle}

\def\be{\begin{equation}}
\def\ee{\end{equation}}

\documentclass[12pt,epsfig]{iopart}
\usepackage{epsfig}
\begin{document}
\setcounter{page}{1}
\title{Phase stability analysis in Fe-Pt and Co-Pt alloy systems: An augmented space study}
\author{\bf Durga Paudyal, Tanusri Saha-Dasgupta and Abhijit Mookerjee}
\address { S.N. Bose National Centre for Basic Sciences,
JD Block, Sector 3, Salt Lake City, Kolkata 700098, India\\
email: dpaudyal@bose.res.in, tanusri@bose.res.in, abhijit@bose.res.in}
\date{\today}

\begin{abstract}
We have studied the problem of phase stability in Fe-Pt and Co-Pt alloy systems.
We have used the orbital peeling technique in the conjunction of augmented space 
recursion based on the tight binding linear orbital method as the method for the 
calculation of pair interaction energies. In particular, we have generalized our 
earlier technique to take into account of magnetic effects for the cases where 
the magnetic transition is higher than the order disorder chemical transition 
temperature as in the case of Co$_3$Pt. Our theoretical results obtained within 
this framework successfully reproduce the experimentally observed trends.
\end{abstract}  

\section{Introduction}
The first-principles study of phase stability in alloy systems has been an active
field of research for past several years. Research over years has established this
field as one of the important stream of research in ab-initio electronic structure
calculations. Theories have been formulated starting from the description of
completely disordered phase \cite{kn:epi} as well as that from ordered super-structures \cite{kn:cw}.  

Alloy systems are in general complicated and modeling of all the relevant effects
active in a particular alloy system is a challenge by itself. The aim is to have a
microscopic understanding and predictive capability. One needs to take into account
systematically the effects like the on-site and off-site disorder, charge-transfer
effect, the effect of local lattice distortions, the short range ordering effect as
is relevant for a given alloy system. For alloys with magnetic component, a further
ingredient, namely magnetism is added to the problem. One might naively think that
one needs to be concerned with magnetism only if one is interested in magnetic
properties of materials. This is however not the case. Rather the formation of stable
ordered structures can depend on properly taking into account magnetism, which can
strongly effect the phase stability. The most illustrative example is that of the
strong ferromagnet Ni-rich Fe-Ni alloys, where the ordering is entirely driven by
magnetism and absence of spin-polarization in calculation leads to wrong ground state
with phase-segregated rather than phase-ordered configuration \cite{kn:ducastelle}. There are also
many other magnetic alloy systems which are not so strong ferromagnets as in the case
of Ni-rich Fe-Ni with completely full majority spin d-states but posses similar
attribute.

In our present study, we considered the problem of phase stability in Fe-Pt and
Co-Pt alloys. There are many studies on magnetic, optical and magneto-optical
characterization in these alloys \cite{kn:uba}. Nevertheless, a systematic first-principles
study on chemical ordering tendency is lacking. The phase diagrams of Fe-Pt \cite{kn:fept} and
Co-Pt \cite{kn:copt} alloy systems show that the magnetic transition temperature is below the
chemical order-disorder transition temperature in most part of the phase diagram
except in the region of high concentration of Co (above 60$\%$) in Co-Pt alloys.
In this region one would therefore expect a strong influence of magnetism on the
chemical order. In our approach, the magnetism part is dealt within Stoner theory with
rigidly exchange-split, spin-polarized band. Within this approach the paramagnetic phase
has ``no exchange-splitting" and the magnetization in the paramagnetic phase is lost via
Stoner particle-hole excitations. However, the other point of view of describing the
paramagnetic phase could be that of the local-moment formation where the average over the
local moment's orientations produce zero overall magnetization but nevertheless
there exists a local-moment disorder \cite{kn:staunton}. Whether such description is necessary or
not depends on the time-scale associated with the rate of change of orientation of
the local moments as compared to the time scale of electronic motion.
Staunton et.\etal \cite{kn:staunton} have shown that consideration of such local moment formation
can be important in describing properly the atomic short-range order data in FeV
system. We are yet to explore the effect of such local moment formations for the
paramagnetic Fe-Pt and Co-Pt alloys. However our results obtained within Stoner approach
already show reasonable agreement with experimental result, suggesting the necessity
of inclusion of such effect in a second level. Our calculational scheme is that of
augmented space recursion implemented within the framework of first-principles
electronic structure calculation of TB-LMTO. Our scheme has been already proved to
be efficient to handle the issues of off-diagonal disorder, large charge transfer
effect, local lattice distortion which are important for alloys with large size
mismatch between components and components with very different valences as is the
case in Fe-Pt and Co-Pt. Due to the presence of the high-mass element Pt, relativistic
effect also turn out to be crucial, which has been dealt within the scalar relativistic
theory. We perform a thorough analysis of the phase stability in terms of pair interaction, 
effective pair potential surfaces, instability temperatures 
and atomic short range order maps.

\section{Formalism}

We start from a completely disordered alloy. Each site $R$ has an occupation
variable $n_R$ associated with it. For a homogeneous perfect disorder
$\langle n_R\rangle = x$, where $x$ is the concentration of one of the
components of the alloy. In this homogeneously disordered system we now
introduce fluctuations in the occupation variable at each site : $\delta x_R =
n_R - x$. Expanding the total energy in this configuration about the
energy of the perfectly disordered state we get :

\begin{equation}
E(x) \eq  E^{(0)}\plus \sum_{R=1}^{N} E_{R}^{(1)}\ \delta x_{R} \plus
\sum_{RR'=1}^{N} E_{RR'}^{(2)}\ \delta x_{R}\ \delta x_{R'} \plus \ldots
\label{eq:eq1}
\end{equation}

\n  The coefficients $E^{(0)}$  , $E_R^{(1)} \ \ldots $  are  the  effective
renormalized cluster interactions.
$E^{(0)}$  is the energy of the averaged disordered
medium. If we embed atoms of the type A or B at $R$ in the disordered
background and the total energies are $E_A$ and $E_B$, then by the above
equation :

\[  E_{R}^{(1)}  \eq E_A \mns E_B\]

\n This  one  body  interaction  results  from  the
interchange of a $B$ atom with an $A$ atom at site $R$  in  the  alloy.
\vskip 0.2cm
\n Similarly, $E_{RR'}^{(2)}$   is the effective renormalized pair interaction
which is the difference in the one body interactions at $R$,
when site $R'$  ($\not= R$) is  occupied
either by  an A  or a B  atom.

\[  E_{RR'}^{(2)} \eq E_{AA}\pls E_{BB}\mns E_{AB} \mns E_{BA} \]

For magnetic pair interaction energy we take the averaged non magnetic disordered medium and embed two same atoms with
two different spins up and down and calculate pair interaction energy as explained above. The same we repeat for other
components with two different spins up and down. Then we embed the different atoms with same as well as two
different spins up and down. This procedure gives us the magnetic pair interaction energies which are given as:

\[J^{(2)}_{AA} = E^{\uparrow \uparrow}_{AA} + E^{\downarrow \downarrow}_{AA} - E^{\uparrow \downarrow}_{AA}
- E^{\downarrow \uparrow}_{AA}\]

Similarly,
\[J^{(2)}_{BB} = E^{\uparrow \uparrow}_{BB} + E^{\downarrow \downarrow}_{BB} - E^{\uparrow \downarrow}_{BB}
- E^{\downarrow \uparrow}_{BB}\]

And,
\[J^{(2)}_{AB} = E^{\uparrow \uparrow}_{AB} + E^{\downarrow \downarrow}_{AB} - E^{\uparrow \downarrow}_{AB}
- E^{\downarrow \uparrow}_{BB}\]

Therefore the effective magnetic pair interaction energy is given as:
\[J^{(2)} = J^{(2)}_{AA} + J^{(2)}_{BB} - 2 J^{(2)}_{AB}\]

Thus we can arrive at the relation of effective pair interaction energy including magnetism as:
\begin{equation}
E_{RR'}^{(2)} = E^{(2)}_{AA} + E^{(2)}_{BB} - 2E^{(2)}_{AB} + J^{(2)}_{AA} + J^{(2)}_{BB} - 2 J^{(2)}_{AB}
\end{equation}

We  will  retain  terms  up  to  pair
interactions  in  the  configuration  energy  expansion. Higher
order interactions may be included for a more accurate and
complete description. For the phase stability study it is the pair interaction which
plays the dominant role.
\vskip 0.2cm
\n The total energy of a solid may be separated into two terms : a
one-electron band contribution $E_{BS}$ and the electrostatic contribution $E_{ES}$.
The renormalized cluster interactions 
should, in principle, include both $E_{BS}$ and $E_{ES}$
contributions. Since the renormalized cluster interactions
involve the difference of cluster energies, it is usually assumed
that the electrostatic terms cancel out and only the band
structure contribution is important. Such an
assumption though is not rigorously true, has been shown to 
hold good in a number of alloy systems \cite{kn:heine}.
\n Considering only band structure contribution, the effective pair 
interactions may be written as :

\begin{equation}
E_{RR'}^{(2)} \eq -\int_{-\infty}^{E_F} dE \left\{  -\frac{1}{\pi} \Im m~\log 
\sum_{IJ}\det \left(G^{IJ}(E)\right)~ \xi_{IJ}\right\} 
\end{equation}

\n where,
 $G^{IJ}$ represents the configurationally averaged Green function corresponding to the disordered Hamiltonian
whose R and R$^\prime$ sites are occupied by I-th and J-th type of atom, and

\[ \xi_{IJ} \eq \left\{\begin{array}{ll}
                         +1  & \mbox{if I=J}\\
                         -1  & \mbox{if I$\not=$ J}
                         \end{array} \right. \]

The behavior of this function is quite complicated and hence the integration by standard routines (e.g.
Simpson's rule or Chebyshev polynomials) is difficult, involving many iterations before convergence is achieved.
Furthermore the integrand is multi-valued, being simply the phase of $\sum_{IJ}\det \left(G^{IJ}\right)~\xi_{IJ}$.
The way out for this was suggested by Burke \cite{kn:op} which relies on the repeated application of the
partition theorem on the Hamiltonian H$^{IJ}$. The final result is given simply in terms
of the zeroes and poles of the Green function in the region $E<E_F$

\begin{equation}
\fl E^{(2)}_{RR'} \eq 2 \sum_{IJ}\xi_{IJ} \sum_{k=0}^{\ell max} \left[ \sum_{j=1}^{z^{k,IJ}}\ Z^{k,IJ}_j 
\mns \sum_{j=1}^{p^{k,IJ}}\ P^{k,IJ}_j \pls \left( p^{k,IJ} - z^{k,IJ}\right)\ E_F\right] 
\end{equation}

where $Z^{k,IJ}_j$ and $P^{k,IJ}_j$ are the zeros and poles of the peeled Green's function $G^{IJ}_{k}$ 
of disordered Hamiltonian with occupancy at sites R and R$^\prime$ by I and J of which first (k-1) rows and 
columns has been deleted. $p^{k,IJ}$ and $z^{k,IJ}$ are the number of poles and  zeroes in the energy region 
below $E_F$. 

The calculation of the effective pair interaction with out magnetism as well as with magnetism taken 
in to account in  our  formalism  reduces  to  the determination of the peeled configuration averaged  
green  functions $\langle$G$^{IJ}_k\rangle $. We employed  the  augmented space recursion coupled 
with the linearized tight-binding  muffin tin orbital method (TB-LMTO) introduced by  Andersen  and  
Jepsen \cite{kn:lmto} for a first principles determination  of  these  configuration averaged quantities.   
We took the  most  localized,  sparse tight  binding  first order Hamiltonian  derived systematically 
from the LMTO theory within the atomic sphere approximation (ASA) and generalized to random alloys.  
The augmented space recursion method with effective Hamiltonian used for recursion in augmented space 
for the calculation of the peeled Green functions has been described in earlier paper \cite{kn:dta}. 
We refer the readers to this paper and references therein for the details. 

For non-isochoric alloys, the difference in atomic radii of the constituents lead to change in the 
electronic density of states. One thus expects that the mismatch of size produces, in addition to a
relaxation energy contribution, a change in the band structure. Within our Augmented Space Recursion 
(ASR), off-diagonal disorder in the structure matrix because of local lattice distortions 
due to size mismatch of the constituents, can be handled on the same footing as diagonal disorder in 
the potential parameters \cite{kn:big,kn:mook}.

The augmented space recursion with the TB-LMTO  Hamiltonian
coupled with orbital peeling allows us to  compute  configuration
averaged  pair interaction energies directly,  without  resorting  to  any
direct averaging over a finite number of  configurations. In an earlier 
communication \cite{kn:ks} we have discussed how one uses the local 
symmetries of the augmented space to reduce the Hamiltonian and carry out 
the recursion on a reducible subspace of much lower rank. If we fix the 
occupation of two sites, the local symmetry of the augmented space is 
lowered (this is very similar to the lowering of spherical symmetry to 
cylindrical symmetry when a preferred direction is introduced in an 
isotropic system). We may then carry out the recursion in a suitably 
reduced subspace.

For the calculation of instability temperature using these pair interaction energies, 
we have used Khachaturian's concentration wave 
approach in which the the stability of a solid solution with respect to a small concentration 
wave of given wave vector ${\bf k}$ is guaranteed as long as $ k_BT + V({\bf k})\ x(1-x) > 0 $.
Instability of the disordered state sets in when :
\begin{equation}
k_B~T^{i} + V({\bf k})\ x(1-x) = 0
\end{equation}
$T^{i}$ is the instability temperature corresponding to a given
concentration wave disturbance.  $V({\bf k})$ is the Fourier transform of pair interaction energies 
and {\it x} is the concentration of one of the constituent atoms. The details are given in our 
previous paper \cite{kn:dta} and references therein. This approach amounts to treating the 
entropy part within the Bragg-Williams mean field theory.

Experimentally the instability of the disordered phase to ordering may be seen in electron, x-ray or neutron 
scattering measurements. These are directly related to the Warren-Cowley short range order parameter $\alpha(\bf k)$ 
which in turn is related to effective pair energies through \cite{kn:staunton1}

\begin{equation}
\alpha({\bf k}) = \frac{\beta~x~(1-x)}{1-\beta~x~(1-x)~V({\bf k})} 
\end{equation}
Where, $\beta = \frac{1}{k_B T}$.

\section{Computational details}
We have performed the total energy density functional calculations for the ab-initio electronic structure 
description of alloys. The Kohn-Sham equations were solved 
in the local density approximation (LDA) for non magnetic calculations and local spin density approximation 
(LSDA) for magnetic calculations with von Barth-Hedin (vBH) \cite{kn:vbh} exchange correlations. The 
calculations have been performed in the basis of tight binding linear muffin-tin orbitals in the
atomic sphere approximation (TB-LMTO-ASA) \cite{kn:lmto} \cite{kn:oka}-\cite{kn:ddsam} including combined corrections. 
The calculations are semi-relativistic through inclusion of mass-velocity and Darwin correction terms. 
The k-space integration was carried out with 32$\times$32$\times$32 mesh resulting 969 k points for cubic
primitive structures in the irreducible part of the 
corresponding Brillouin zone. The convergence of the total energies with respect to k-points have been checked. 
To have theoretical estimates of the equilibrium lattice parameters, we have carried out the minimization of the 
self-consistent TB-LMTO-ASA total energies varying lattice parameters for Fe-Pt and Co-Pt alloys at different 
concentrations. In Table 1, we have quoted thus obtained equilibrium lattice parameters that were used to calculate the 
self consistent potential parameters which were then used to calculate the pair interaction energies. For 25$\%$ 
concentration of Pt we have also calculated the equilibrium lattice parameter with magnetic contribution which 
has better agreement with corresponding experimentally predicted value. In all cases we have obtained
lower equilibrium lattice parameters as compared to experimental ones. This is characteristic of the local density
approximation which overestimates bonding.

\begin{table}[t]
\caption{The theoretically obtained equilibrium lattice constants from non magnetic calculations as compared to experimental
lattice constants. The values inside the bracket in the case of 25$\%$ concentration of Pt in Co-Pt are with the magnetic contribution}
\begin{center}
\begin{tabular}{ccc}
\hline\hline
concentration & Calculated lattice & Experimental lattice \\
of Pt         & parameter in a.u.    & parameter in a.u.      \\ \hline \\
\multicolumn{3}{c}{\bf Fe$_{1-x}$Pt$_{x}$} \\ \\
0.25                &  6.72                              & 7.05\\
0.50                &  7.05                              & 7.25\\
0.75                &  7.18                              & 7.31\\ \\
\multicolumn{3}{c}{\bf Co$_{1-x}$Pt$_{x}$} \\ \\
0.25                &  6.71[6.78]                        & 6.92\\
0.50                &  7.08                              & 7.20\\
0.75                &  7.18                              & 7.24\\ \hline
\hline
\end{tabular}
\end{center}
\end{table}

The calculation of Madelung potential is a challenging job for disordered alloys due to the absence of lattice 
periodicity. For the treatment of the Madelung potential, we followed the procedure suggested by 
Kudrnovsk\'y \etal \cite{kn:kd}
and use an extension of the procedure proposed by Andersen \etal \cite{kn:lmto}. We have chosen the atomic sphere radii 
of the components in such a way that they preserve the total volume on the average and the individual atomic spheres 
are almost charge neutral. This ensures that total charge is conserved,  but each atomic sphere carries no excess
charge. In doing so, one needs to be careful about the sphere overlap which should be under certain limit so as to 
not violate the atomic sphere approximation.

The effective pair interaction energies are calculated at the Fermi level so one needs to be very careful 
about the convergence of our procedure. 
In fact, errors can arise 
in the augmented space recursion because one can carry out only finite number of recursion steps and then 
terminate the continued fraction using available terminators. Also one chooses a large but finite part of 
the augmented space nearest neighbour map and ignores the part of the augmented space very far from the 
starting state. 

For finding out the Fermi energy accurately, we have used the energy dependent formulation 
of augmented space recursion in which the disordered Hamiltonian with diagonal as well as 
off-diagonal disorder is recast into an energy dependent Hamiltonian having only diagonal 
disorder. This allows one to sample more shells in the augmented space. Though this formulation 
reduces the computational burden, the recursion becomes energy dependent and it is not suitable 
to carry out one recursion per energy point. This is tackled by choosing a few seed points 
across the energy spectrum uniformly and then carry out recursion on those points and spline fit 
the coefficients of recursion through out the whole spectrum. This enabled us to carry out large 
number of recursion steps since the configuration space grows significantly less faster for diagonal 
as compared to off diagonal disorder.
Convergence of physical quantities with recursion steps have been
discussed in detail earlier by Ghosh \cite{kn:sdgthesis}.

We have checked  the convergence of pair interaction energies with respect to recursion steps. 
The convergence of Fermi energy has also been checked with respect to recursion steps and 
seed energy points. We have found that the pair interaction energies and Fermi energy converge 
beyond seven recursion steps. We also found that we need at least thirty five seed energy 
points to get the convergence of 
Fermi energy. Our calculations reported in the following have been guided by these convergence 
properties.

\section{Results and discussions}

\subsection{Pair interaction energies}

\begin{table}
\caption{Pair interaction energies upto fourth nearest neighbour in mRyd/atom. The values inside bracket in the
case of 25$\%$ concentration of Pt in Co-Pt are with the magnetic contribution.}
\begin{center}
\begin{tabular}{cccccc}
\hline
Concentration & V$_1$ & V$_2$ & V$_3$ & V$_4$ \\
of Pt (x)     &       &       &       &       \\  \hline \\

\multicolumn{5}{c}{\bf Fe$_{1-x}$Pt$_{x}$}\\ \\

0.25          & 5.98[4.73]  & -0.12[-0.15] & 0.39[0.15] & -0.15[-0.33] \\
0.50          & 5.80  & -0.15 & 0.28  & -0.48 \\
0.75          & 4.14  & -0.03 & 0.20  & -0.13 \\ \\

\multicolumn{5}{c}{\bf Co$_{1-x}$Pt$_{x}$}\\ \\

0.25          &  9.97[8.00] & -0.11[-0.13] &  0.24[0.17] & -0.30[-0.21] \\
Capitan \etal \cite{kn:capitan}&  [4.06]     &  [-1.83]     &    [0.39]   & [-0.22] \\\\
0.50          &  8.04 & -0.01 &  0.15 & -0.12 \\
0.75          &  7.07 &  0.08 &  0.10 & -0.18 \\
Sanchez \etal \cite{kn:sanchez}&  1.11 & -2.27 &  0.33 & -0.93 \\
\hline \hline
\end{tabular}
\end{center}
\end{table}

\begin{figure}
\centering
\psfig{file=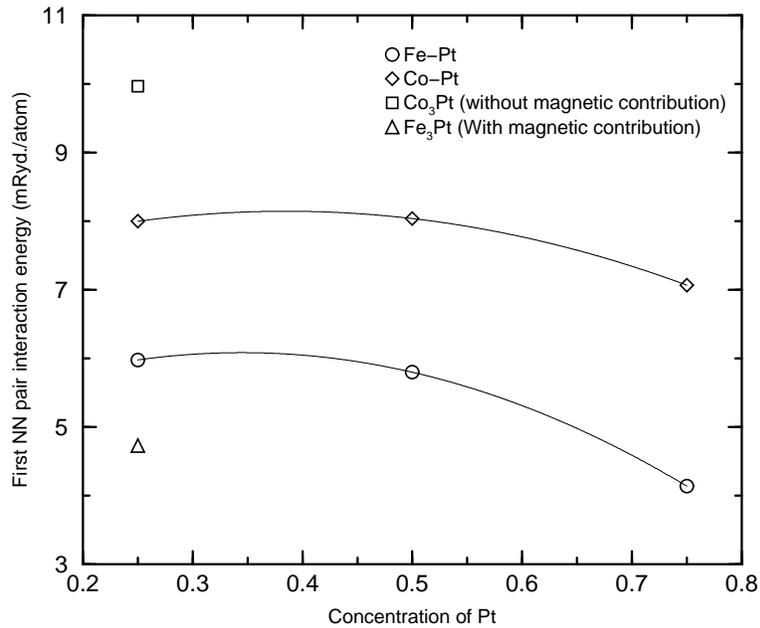,height=10cm,angle=-90}
\caption{Plot for first nearest neighbour pair interaction energies (V$_1$) in Fe-Pt and Co-Pt alloys which shows 
the concentration dependence in the chemical order.}
\end{figure}

The pair interaction energies calculated for 25, 50 and 75$\%$ concentration of Pt in Fe-Pt and Co-Pt  
alloy systems using above explained formalism are shown in the Table 2. In all cases we have obtained positive 
first nearest neighbour pair interaction energies which indicate the ordering tendency in these alloys as par
with the experimentally predicted ordering tendencies. 

From the Figure 1 we see the increase in first nearest neighbour pair interaction energies when one goes
from Fe-Pt to Co-Pt. Figure 1 also shows the strong concentration dependence of pair interaction. 

For 25$\%$ concentration of Pt in Co-Pt alloys, we have included the effect of magnetism since the Curie 
temperature lies above the order disorder transition temperature in the experimental phase diagram \cite{kn:copt}. 
The first nearest neighbour pair interaction energy calculated with the inclusion of magnetic contribution 
comes out to be lower than the corresponding value obtained without magnetic contribution.

In the experimental phase diagram of Fe-Pt alloy system \cite{kn:fept}, the magnetic transition temperature 
(Curie temperature) for Fe$_3$Pt lies much below the order disorder chemical transition temperature. 
So it is not necessary to include the magnetic effect to study the chemical ordering problem.  However to 
compare it with the Co$_3$Pt case we have carried out the calculation for Fe$_3$Pt alloy too. 
Our calculation 
shows the same trend as Co$_3$Pt, namely the dominant V$_1$ interaction decreases on including magnetic effect. 

In Figure 2, we show the density of states with and without magnetic contribution for 25$\%$ concentration of Pt in 
Co-Pt system. In magnetic case we see that the majority spin band almost full. The contribution to the pair 
interaction energy in this case mainly comes from the partially filled band. There is little decrease in the Fermi 
energy in magnetic case as 
compared to non magnetic case. This shifting of Fermi energy slightly reduces the pair interaction energy.      
The low value of first neighbour pair interaction energy in magnetic case is primarily due to the negative 
value of magnetic 
pair interaction energy between two Co atoms.  

In Figure 3, taking the example of 25$\%$ concentration of Pt in Co-Pt with magnetic contribution, we show the variation 
of pair interaction energy as a function of nearest neighbour shells. The pair interaction energy decreases rapidly while 
going from first nearest neighbour to second nearest neighbour shell. For comparison we have also plotted the pair 
interaction energies extracted by Capitan \etal \cite{kn:capitan} from their experimentally measured short 
range order using inverse cluster variation method. Our calculated first and second nearest neighbour pair 
interaction energies are higher than that of Capitan \etal \cite{kn:capitan}. But 
there is agreement in the third and fourth nearest neighbour pair interaction energies. Though there are 
differences in the 
first and second nearest neighbour pair interaction energies between Capitan \etal's \cite{kn:capitan} and ours, 
the trend of pair interaction 
energy as a function of nearest neighbour shells is similar as is seen from the Figure 3.         

\begin{figure}
\centering
\psfig{file=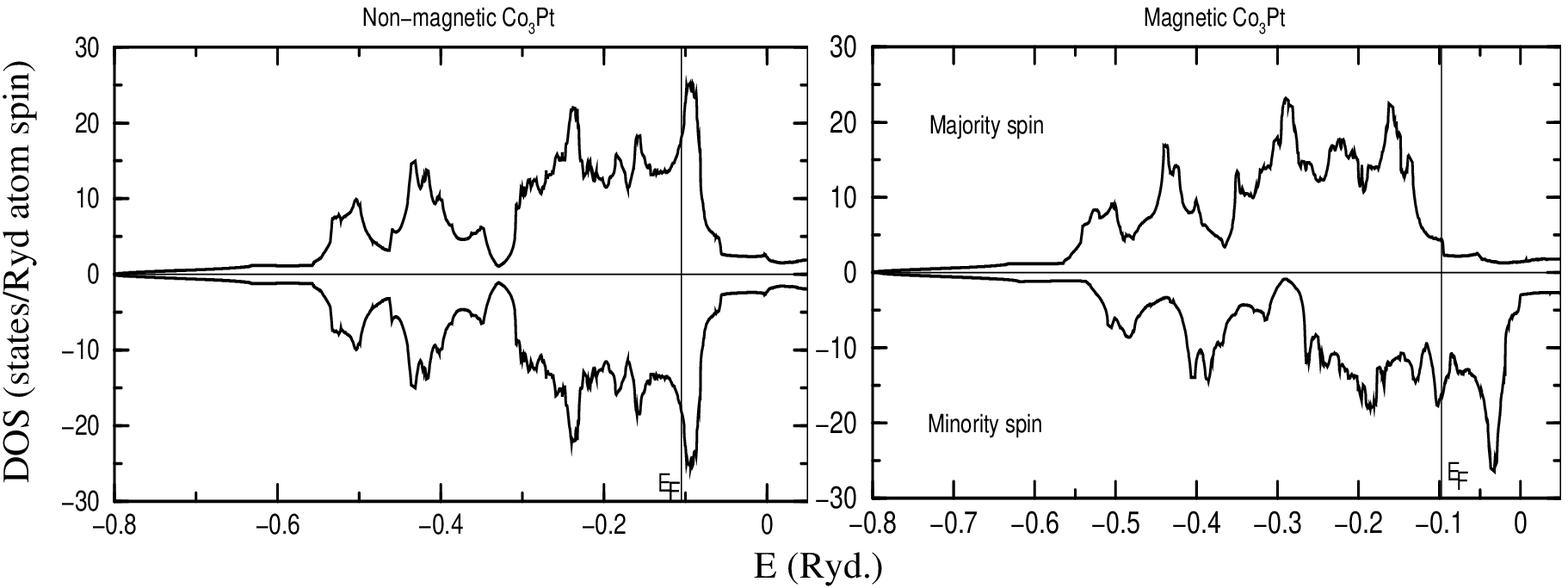,height=5.5cm,angle=0}
\psfig{file=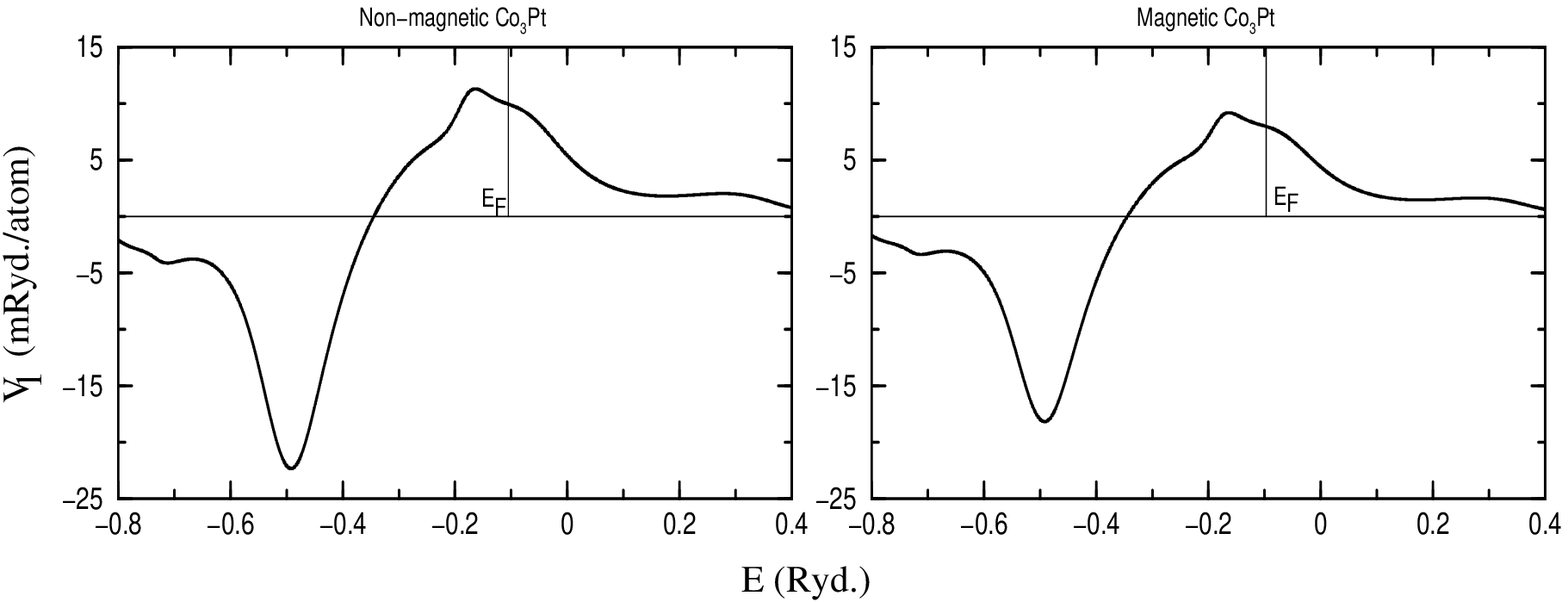,height=5.75cm,angle=0}
\caption{Density of states and first nearest neighbour pair interaction energy as a function of band energy 
in Co$_3$Pt alloy system}
\end{figure}

The pair interaction energies extracted by Sanchez \etal \cite{kn:sanchez} from their experiment on 
short range order using inverse cluster variation method for 75$\%$ concentration of Pt in Co-Pt 
are tabulated in Table 2. Their first neighbour 
pair energy is lower in magnitude than that of second neighbour 
pair interaction energy. Our calculated pair interaction energies instead follow usual trend of 
having higher magnitude of 
first neighbour pair energy than that of second neighbour pair energy. Though there is 
substantial difference in the trend 
of pair interaction energies, the instability temperatures computed using both sets of 
pair interaction energies turn out to be almost same as shown 
in the section 4.4.

\begin{figure}
\centering
\psfig{file=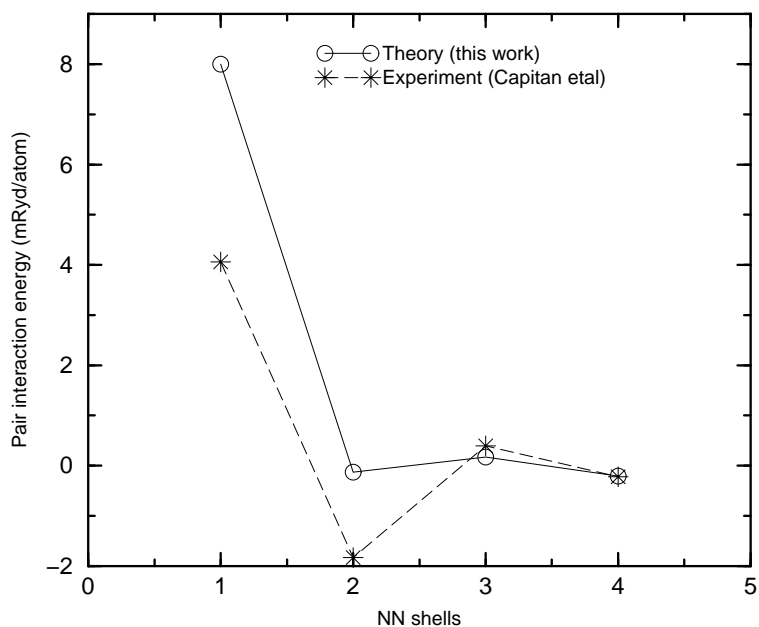,height=10cm,angle=-90}
\caption{The calculated pair interaction energies as a function of shell numbers in Co$_3$Pt as compared to 
experimental estimates of Capitan \etal \cite{kn:capitan}}
\end{figure}

\subsection{Effective pair potential surfaces}

\begin{table}[h]
\caption{Effective potentials V$<{\bf hkl}>$. The values
inside bracket in the case of 25$\%$ concentration of Pt in Co-Pt are with the magnetic contribution.}
\begin{center}
\begin{tabular}{cccccc}
\hline
Concentration & Experimental & V$<{\bf 000}>$ & V$<{\bf 100}>$ & V$<{\bf 1 {\frac{1}{2}}0}>$
& V$<{\bf {\frac{1}{2}}{\frac{1}{2}}{\frac{1}{2}}}>$  \\
of Pt (x)     & ordering     &                &  & &\\ \hline \\
\multicolumn{6}{c}{\bf Fe$_{1-x}$Pt$_{x}$}\\ \\
0.25 & $<{\bf 100}>$ L1$_2$ & 78.60 & -29.56 & -20.44 & -1.08\\
0.50 & $<{\bf 100}>$ L1$_0$ & 69.66 & -32.10 & -19.34 & -4.86\\
0.75 & $<{\bf 100}>$ L1$_2$ & 52.74 & -19.90 & -14.5  & -1.38\\ \\
\multicolumn{6}{c}{\bf Co$_{1-x}$Pt$_{x}$} \\ \\
0.25 & $<{\bf 100}>$ L1$_2$ & 121.14[96.78] &-46.06[-36.66] & -36.98[-30.06] & -2.94[-1.74] \\
Capitan \etal \cite{kn:capitan} & $<{\bf 100}>$ L1$_2$ & [44.46] & [-32.98]   & [-15.90]       & [8.34] \\ \\
0.50 & $<{\bf 100}>$ L1$_0$ & 98.58         & -34.86        & -30.5          & -1.38 \\
0.75 & $<{\bf 100}>$ L1$_2$ & 85.56         & -30.76        & -26.60         & -2.64 \\ \\
Sanchez \etal \cite{kn:sanchez} & $<{\bf 100}>$ L1$_2$ & -3.54 & -31.86       & -2.62          &  2.46 \\
\hline
\end{tabular}
\end{center}
\end{table}

The effective pair potentials $V(\bf{k})$ calculated using pair interaction energies (T = 0) for Fe-Pt and  Co-Pt
alloys at 25, 50 and 75$\%$ concentration of Pt are shown in the Table 3. These effective pair potentials were obtained by 
the Fourier transform of above explained first four nearest neighbour pair interaction energies. The values of 
$V({\bf k})$ for different ordering stars are compared in the Table 3. From the Table the minima are seen at the 
position ${\bf<100>}$. These minima clearly show the L1$_2$ chemical ordering for 25 and 75$\%$ and L1$_0$ chemical 
ordering for 50$\%$ concentration of Pt in these alloy systems. The value for $V(\bf{100})$ increases when one goes from 
25 to 50$\%$ and then decreases while going from 50 to 75$\%$ concentration of Pt in Fe-Pt alloys. In Co-Pt alloys 
The value for $V(\bf{100})$ systematically decreases while going from 25 to 50 and then to 75$\%$ concentration of Pt.     

In 25 $\%$ concentration of Pt in Co-Pt alloy, the magnetic transition temperature is higher than the order
disorder transition temperature. Therefore we have included the effect of magnetism in the chemical order
in this particular alloy. The comparison of V$(\bf{k})$ minima shown in the Figure 4 with magnetic contribution
indeed matches with the
minima obtained using the pair interaction energies extracted from the experimentally measured short range order
parameters using inverse cluster variation method by Capitan \etal \cite{kn:capitan}. This experimental
measurement was done in ferromagnetic state. If we compare with the non-magnetic V$(\bf{100})$ surfaces we see
that there is decrease in the chemical ordering which indicates that the magnetism actually reduces
the chemical ordering in this alloy allowing to order in lower temperature.  
As a comparison the ``artificial" magnetic V${\bf (100)}$ surface of Fe$_3$Pt alloy system shows that taking 
into account the effect of V$_n$ farther than first nearest neighbour, magnetism increases slightly the tendency 
to chemical order over that of the non magnetic case.   

The V$(\bf{100})$ minima obtained for CoPt$_3$ using our theoretically calculated pair interaction energies and that of
extracted pair interaction energies from the experimentally measured short range order parameters using inverse cluster
variation method by Sanchez \etal \cite{kn:sanchez} match well which is shown in Figure 5. The differences seen in the patterns of V$(\bf{k})$
is due to the differences in individual pair interaction energies in real space. Though there is difference in $V({\bf k})$ surface patterns,
the value of $V({\bf 100})$ minima in both cases are very similar.

\begin{figure}
\centering
\psfig{file=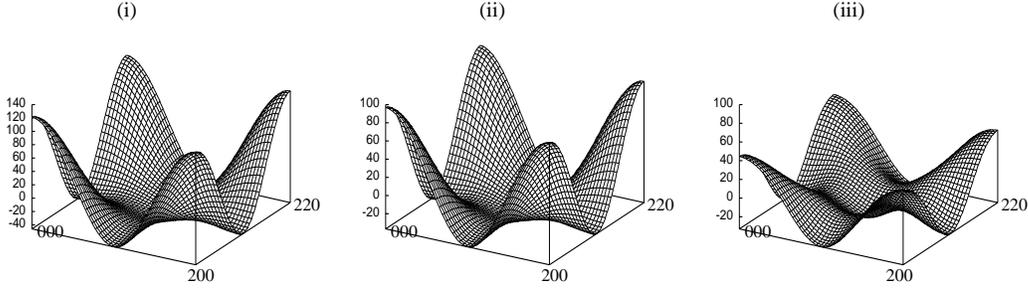,height=4cm,angle=0}
\caption{Effective pair potential V$(\bf{k})$ surfaces for Co$_3$Pt alloy system in the ({\bf hk0}) plane using our theoretically
calculated pair interaction energies (i) without magnetic contribution and (ii) with magnetic contribution and (iii) using extracted 
pair interaction energies by Capitan \etal \cite{kn:capitan} from their experiment.}
\end{figure}

\begin{figure}
\centering
\psfig{file=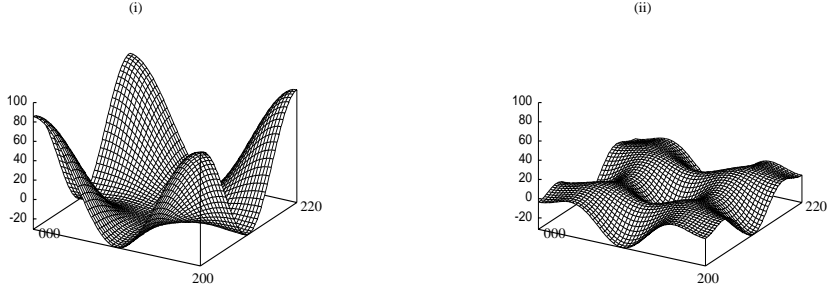,height=4cm,angle=0}
\caption{Effective pair potential V$(\bf{k})$ surfaces for CoPt$_3$ alloy system in the ({\bf hk0}) plane using (i) our 
theoretically calculated pair interaction energies and (ii) extracted pair interaction energies by Sanchez \etal \cite{kn:sanchez} from 
their experiment.}
\end{figure}

\subsection{Instability temperatures}

Using the pair interaction energies obtained by us, we have calculated the instability temperatures 
in Fe-Pt and Co-Pt alloys within Khachaturyan's concentration wave approach as explained in Section 2. 
The results are shown in Figure 6. In Fe-Pt system the calculated order 
disorder transition temperatures are lower than that of corresponding experimental order 
disorder transition temperatures. We however clearly see the correct trend of calculated instability 
temperature with experimentally predicted transition temperature. 

Our calculation with the inclusion of effect of magnetism for 25$\%$ concentration of Pt 
in Co-Pt alloys shows the instability 
temperature closer to experimental order disorder transition temperature than that of the 
calculated instability temperature with out magnetic 
effect. This is also in good agreement with the value obtained using the pair interaction energies 
extracted from the experimentally 
measured short range order parameters using inverse cluster variation method by Capitan \etal \cite{kn:capitan}. 
For 50$\%$ concentration of Pt 
our calculated instability temperature matches with the calculation by Staunton's group \cite{kn:staunton2} using KKR-CPA based method. 
The calculated values of instability temperature are much higher than the experimentally predicted order disorder transition temperatures. The values of instability  
temperatures obtained by us and Staunton's group \cite{kn:staunton2} and that of Sanchez's 
group \cite{kn:sanchez} (using extracted pair 
interaction energies upto fourth nearest neighbour from the experimentally measured short range order parameters 
using inverse cluster variation method) for 75$\%$ concentration of Pt are almost similar and slightly 
lower than the experimental predictions of order disorder transition temperatures.    

\begin{figure}
\centering
\psfig{file=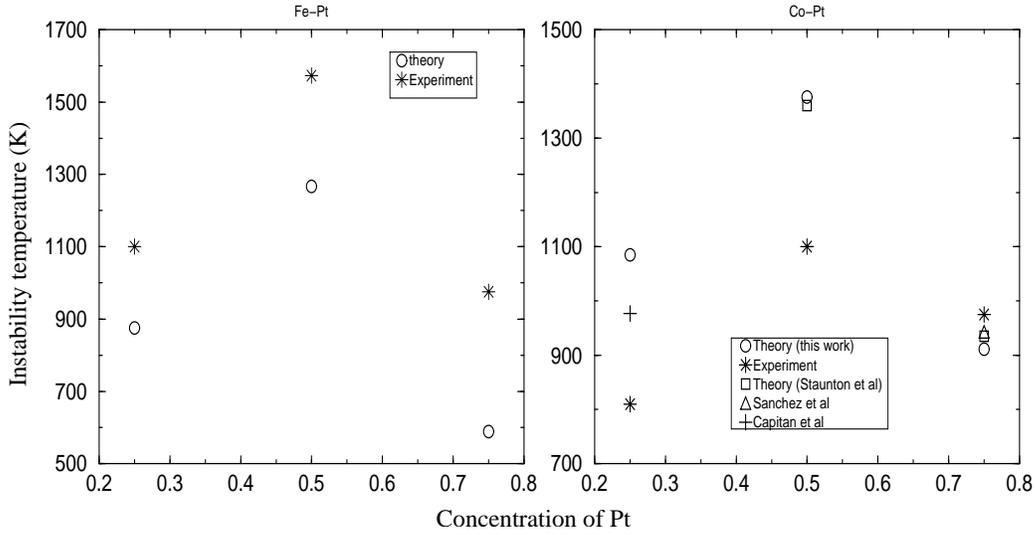,height=7cm,angle=0}
\caption{Theoretically calculated instability temperatures as compared to experimental findings
of order disorder transition temperatures.}
\end{figure}

The experimental phase diagram of Co-Pt shows that the order disorder transition temperature for 75$\%$ 
concentration of Pt higher than 25$\%$ concentration of Pt. But the experimental phase diagrams of Fe-Pt 
and Ni-Pt show the order disorder transition temperature for 25$\%$ concentration of Pt higher than 75$\%$ 
concentration of Pt. Our calculation for Fe-Pt and previous calculation for Ni-Pt \cite{kn:dta} show the 
similar trend of instability temperature as that of experimental trend of order disorder transition 
temperature. In Co-Pt our calculated instability temperature at 25$\%$ concentration of Pt is higher than that 
of 75$\%$ concentration of Pt. But the experimental order disorder transition temperature is otherwayround as 
pointed out above. This calls the need for further investigation in terms of important effects that 
might have been overlooked. One important effect among these could have been the neglect of possible local 
moment formations in the paramagnetic CoPt and CoPt$_3$ alloy systems which might be relatively more 
important than that in Fe-Pt systems. The next important effect to be taken into account in general for all 
alloy systems studied is the effect of electrostatic contribution in the effective pair interaction energies 
and the effect of multisite interactions in addition to the pair interaction.  

Our comparison of calculated transition temperatures for Fe-Pt and Co-Pt shows that the 
transition temperatures for corresponding concentrations of Pt in these alloys increase as we go from 
Fe-Pt to Co-Pt (and then to Ni-Pt \cite{kn:dta}). 

\subsection{Short range order}

The SRO parameters $\alpha({\bf k})$ for different ordering stars calculated using pair interaction energies (T = 0) 
for Fe-Pt and Co-Pt alloys at 25, 50 and 75$\%$ concentration of Pt are shown in the Table 4. These SRO values 
were calculated at 10 K above the instability temperatures using above 
explained first four nearest neighbour pair interaction energies to see the effect of SRO in the disordered phase. 
These SRO values show the peak positions 
at {$\bf<100>$}. These peak positions correspond to the diffused scattering peaks which clearly show 
the L1$_2$ short range ordering for 25 and 75$\%$ and L1$_0$ short range ordering for 50$\%$ concentration 
of Pt in the disordered phase of these alloy systems. In Figure 7 we show the concentration dependence of SRO peaks 
$\alpha<{\bf 100}>$. For 50$\%$ concentration of Pt in Fe-Pt and Co-Pt the magnitude of SRO peak is maximum. Next 
higher peak magnitude  is for 25$\%$ concentration of Pt in these alloys and the least is for 75$\%$ concentration of Pt. 
The trend matches with the trend of experimental order disorder transition temperature.

\begin{table}[h]
\caption{Short range order $\alpha<{\bf hkl}>$. The values
inside bracket in the case of 25$\%$ concentration of Pt in Co-Pt are with the magnetic contribution.}
\begin{center}
\begin{tabular}{cccccc}
\hline
Concentration & Experimental & $\alpha<{\bf 000}>$ & $\alpha<{\bf 100}>$ & $\alpha<{\bf 1 {\frac{1}{2}}0}>$ 
& $\alpha<{\bf {\frac{1}{2}}{\frac{1}{2}}{\frac{1}{2}}}>$\\
of Pt (x)     & ordering     & & & &                    \\  \hline \\
\multicolumn{6}{c}{\bf Fe$_{1-x}$Pt$_{x}$}\\ \\
0.25 & $<{\bf 100}>$ L1$_2$ & 0.009 & 2.987 & 0.106 & 0.035 \\
0.50 & $<{\bf 100}>$ L1$_0$ & 0.010 & 3.965 & 0.077 & 0.036 \\
0.75 & $<{\bf 100}>$ L1$_2$ & 0.014 & 2.720 & 0.173 & 0.053 \\ \\
\multicolumn{6}{c}{\bf Co$_{1-x}$Pt$_{x}$} \\ \\
0.25 & $<{\bf 100}>$ L1$_2$ & 0.006[0.004] & 2.834[3.045]   & 0.106[0.144] & 0.023[0.028]\\
Capitan \etal \cite{kn:capitan} & $<{\bf 100}>$ L1$_2$ & [0.013]         & [3.063]      &      [0.057] & [0.024] \\ \\   
0.50 & $<{\bf 100}>$ L1$_0$ & 0.007        & 3.943  & 0.217 & 0.030 \\
0.75 & $<{\bf 100}>$ L1$_2$ & 0.009        & 2.851  & 0.222 & 0.035 \\ 
Sanchez \etal \cite{kn:sanchez} & $<{\bf 100}>$ L1$_2$ & 0.035 & 3.014  & 0.034 & 0.025 \\
\hline
\end{tabular}
\end{center}
\end{table}

The SRO patterns for Co$_3$Pt with magnetic contribution match with the patterns obtained by the Fourier transform 
of Capitan \etal's \cite{kn:capitan} experimental real space SRO parameters which is shown in the Figure 8. 
If we compare with the non-magnetic SRO ${\bf<100>}$ 
patterns we see there is enhancement in the SRO peak which indicates that the magnetism actually enhances the short 
range ordering in the disordered phase of this alloy.  

The SRO patterns obtained for CoPt$_3$ using our theoretically calculated pair interaction energies and that the patterns 
obtained by the Fourier transform of Sanchez \etal's \cite{kn:sanchez} real space SRO parameters agree reasonably 
well, although the effective pair potential surfaces have been shown to differ quite a bit as discussed in section 4.3. 
This in turn points to nonuniqueness of the scheme to extract the pair interaction energies from experimentally 
measured SRO data.

\begin{figure}
\centering
\psfig{file=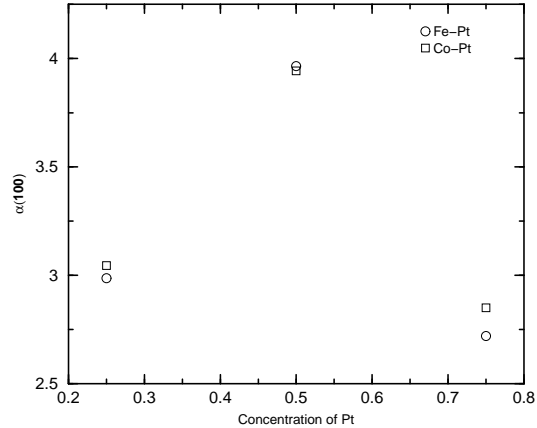,height=7cm,angle=-90}
\caption{Plot for SRO $\alpha$({\bf 100}) values in Fe-Pt and Co-Pt alloys which shows the concentration dependence in
the short range order in the disordered phase of these alloy systems.}
\end{figure}

\begin{figure}
\centering
\psfig{file=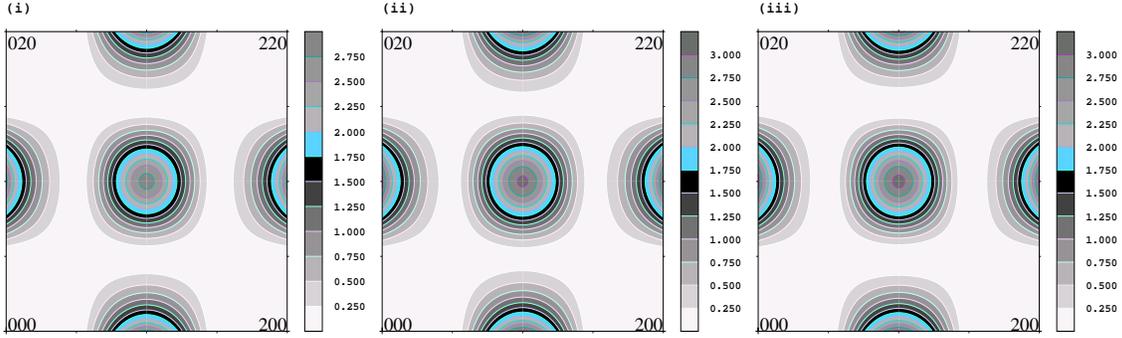,height=5cm,angle=0}
\caption{SRO ($\alpha(\bf{k})$) patterns for Co$_3$Pt alloy system in the ({\bf hk0}) plane using our theoretically
calculated pair interaction energies (i) without magnetic contribution and (ii) with magnetic contribution and (iii) 
using real space SRO parameters from the ferromagnetic experimental measurement of Capitan \etal \cite{kn:capitan}. The peaks in the contour 
plots locate the peaks in the short range order patterns. The plots were drawn at 10 K above the calculated instability 
temperature.}
\end{figure}

\begin{figure}
\centering
\psfig{file=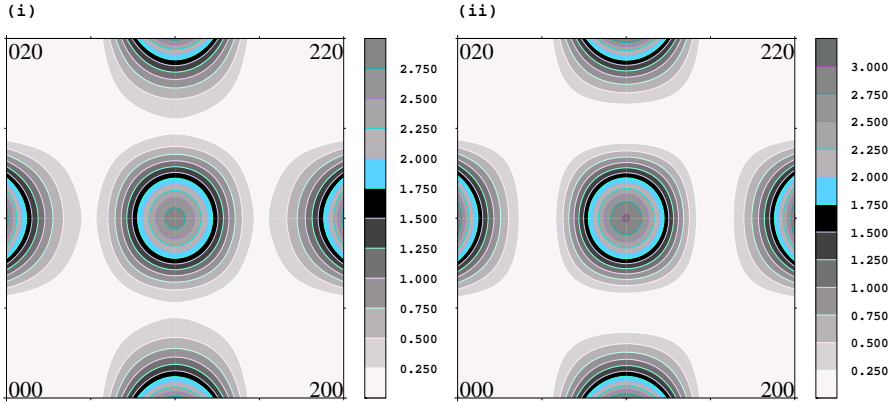,height=6cm,angle=0}
\caption{SRO ($\alpha(\bf{k})$) patterns for CoPt$_3$ alloy system in the ({\bf hk0}) plane using (i) our theoretically 
calculated pair interaction energies and (ii) experimentally measured real space SRO values by Sanchez \etal \cite{kn:sanchez}. 
The peaks in the contour plots locate the peaks in the short range order patterns. The plots were drawn at 10 K above the 
calculated instability temperature.}
\end{figure}
 
\section{Conclusion} 
We have applied our theory for chemical order in metallic alloys for Fe-Pt and Co-Pt systems. Our investigation 
indicates the chemical ordering tendency in these alloys. There is short range ordering tendency in the disordered 
phase of these alloys. Taking the example of Co$_3$Pt we have demonstrated how the magnetism plays a role in 
chemical order within the Stoner approach. We have compared our calculations of atomic short range order in Co$_3$Pt 
and CoPt$_3$ with diffuse x-ray and neutron scattering experiments and obtained fair agreements. In this study of 
Fe-Pt and Co-Pt alloys we have demonstrated that our augmented space recursion method coupled with orbital peeling 
in the basis of linear muffin tin orbital is capable for accurate prediction of chemical order. This method allows 
one to take proper account of charge transfer effect, off-diagonal disorder effect and local lattice distortion 
which are important for alloys with large size mismatch between the components which may not be fully taken into 
account in the mean field based theories like single site coherent potential approximation.


\section*{References}

\begin{thebibliography}{99}
\bibitem{kn:epi} Gonis A., Zhang X. G., Freeman A. J., Turchi P.,
Stocks G. M.  and Nicholson D. M., 1987 \PR  {\bf B36} 4630 (1987)
\bibitem{kn:cw} Connolly J. W. D. and Williams A. R., 1983 \PR {\bf B27} 5169.
\bibitem{kn:ducastelle}Ducastelle F., Order and Phase Stability in Alloys (Elsevier Science, New York, 1991). 
\bibitem{kn:uba} Uba S. {\it et al} 1998 \PR {\bf B57} 1534; Geerts W. 1994  {\it et al} \PR {\bf B50} 12581; Weller D., Harp G.R.,
Farrow R.F.C., Cebollada A. and Sticht J. 1994 \PRL {\bf 72} 2097
\bibitem{kn:fept}Stahl B., Ellrich J., Theissmann R., Ghafari M., Bhattacharya S., Hahn H., Gajbhiye N. S., Kramer D., 
Viswanath R. N., Weissmüller J. and Gleiter H. 2003 Phys. Rev. B {\bf67}, 014422
\bibitem{kn:copt}Sanchez J. M., Moran-Lopez J. L., Leroux C. and Cadeville M.C. 1988 J. Phys. C: Solid State Phys. 
{\bf21} L1091
\bibitem{kn:staunton}Staunton J. B., Ling M. F. and Johnson D. D. 1997 J. Phys.: Condens. Matter {\bf9} 1281
\bibitem{kn:heine} Heine V., (1988) {\it Solid State Physics} {\bf 35} (Academic Press, N. Y. ) 
\bibitem{kn:op} Burke N. R., 1976 {\it Surf.  Sci.}  {\bf 58} 349 
\bibitem{kn:lmto} Andersen O. K.  and Jepsen O., 1984 \PRL {\bf 53} 2571 
\bibitem{kn:dta} Paudyal D., Saha-Dasgupta T. and Mookerjee  A., 2003 \JPCM {\bf 15} 1029
\bibitem{kn:big}  Saha T. , Dasgupta I.  and Mookerjee A. , 1995 \PR {\bf B51} 3413 
\bibitem{kn:mook} Saha T.,  Dasgupta I.  and Mookerjee A., 1996 J.  Phys.  Condens.  Matter.  {\bf 8} 2915
\bibitem{kn:ks} Saha K.K., Saha-Dasgupta T., Mookerjee A. and Dasgupta I., 2004
J. Phys.: Condens. Matter {\bf16} 1409
\bibitem{kn:staunton1} Ling M. F. , Staunton J. B. and Johnson D. D. 1994 J. Phys.: Condens. Matter {\bf6} 5981
\bibitem{kn:vbh} von Barth U. and Hedin L., 1972 \JPC {\bf 5} 1629
\bibitem{kn:oka} Andersen O.K., Jepsen O. and \v{S}ob M., {\sl Electronic Band Structure and Its Applications}
  ed. M. Yussouff. Lecture Notes in Physics 283, Springer 1987 , 1992.
\bibitem{kn:oja} Andersen O.K., Jepsen O. and Krier G., {\sl Lectures on Methods of
Electronic Structure Calculations} eds.: V. Kumar, O. K. Andersen, A. Mookerjee. Singapore, World
  Scientific , 1994.
\bibitem{kn:ddsam} Das G.P., {\sl Electronic Structure of Alloys, Surfaces and Clusters}, Advances in
Condensed Matter Science, Vol. {\bf 4,} eds.: A. Mookerjee and D. D. Sharma, Taylor-Francis, 2003
\bibitem{kn:kd} Kudrnovsk\'y J. and Drchal V., 1990 \PR {\bf B41} 7515
\bibitem{kn:sdgthesis} Ghosh S.D., 2000 Ph.D. Thesis, Jadavpur University
\bibitem{kn:capitan}Capitan M., Lefebvre S., Calvayrac Y., Bessière M., and Cénédèse P., 1999 J. Appl. Crystallogr. {\bf32}, 1039 
\bibitem{kn:sanchez}Kentzinger E., Parasote V., Pierron-Bohnes V., Lami J. F., Cadeville M. C., Sanchez J. M., Caudron R., 
and Beuneu B. 2000 Phys. Rev. B {\bf61}, 14975
\bibitem{kn:staunton2} Razee S. S. A., Staunton J. B., Ginatempo B., Bruno E., and Pinski F. J. 2001 Phys. Rev. B{\bf64}, 014411
\end{thebibliography}
\end{document}